# Learning-based Single-step Quantitative Susceptibility Mapping Reconstruction Without Brain Extraction

[1,2] Hongjiang Wei, [2] Steven Cao, [3] Yuyao Zhang, [4] Xiaojun Guan, [5] Fuhua Yan, [6] Kristen W. Yeom, [2,7] Chunlei Liu

[1] Institute for Medical Imaging Technology, School of Biomedical Engineering, Shanghai Jiao Tong University, Shanghai, China;

[2] Department of Electrical Engineering and Computer Sciences, University of California, Berkeley, CA, USA;

[3] School of Information and Science and Technology, ShanghaiTech University, Shanghai, China

[4] Department of Radiology, The Second Affiliated Hospital, Zhejiang University School of Medicine, Hangzhou, China;

[5] Department of Radiology, Rui Jin Hospital, Shanghai Jiao Tong University, School of Medicine, Shanghai, China;

[6] Department of Radiology, Lucile Packard Children's Hospital, Stanford University, Palo Alto, California, USA.

[7] Helen Wills Neuroscience Institute, University of California, Berkeley, CA, USA.

**Running Title:** Automatic QSM reconstruction

**Correspondence Address:**

Chunlei Liu, Ph.D.

Department of Electrical Engineering and Computer Sciences, and the Helen Wills Neuroscience Institute at the University of California, Berkeley.

505 Cory Hall, Berkeley, CA 94720, USA.

Email: chunlei.liu@berkeley.edu

**Abstract**

Quantitative susceptibility mapping (QSM) estimates the underlying tissue magnetic susceptibility from MRI gradient-echo phase signal and typically requires several processing steps. These steps involve phase unwrapping, brain volume extraction, background phase removal and solving an ill-posed inverse problem relating the tissue phase to the underlying susceptibility distribution. The resulting susceptibility map is known to suffer from inaccuracy near the edges of the brain tissues, in part due to imperfect brain extraction, edge erosion of the brain tissue and the lack of phase measurement outside the brain. This inaccuracy has thus hindered the application of QSM for measuring susceptibility of tissues near the brain edges, e.g., quantifying cortical layers and generating superficial venography. To address these challenges, we propose a learning-based QSM reconstruction method that directly estimates the magnetic susceptibility from total phase images without the need for brain extraction and background phase removal, referred to as autoQSM. The neural network has a modified U-net structure and is trained using QSM maps computed by a two-step QSM method. 209 healthy subjects with ages ranging from 11 to 82 years were employed for patch-wise network training. The network was validated on data dissimilar to the training data, e.g. *in vivo* mouse brain data and brains with lesions, which suggests that the network generalized and learned the underlying mathematical relationship between magnetic field perturbation and magnetic susceptibility. Quantitative and qualitative comparisons were performed between autoQSM and other two-step QSM methods. AutoQSM was able to recover magnetic susceptibility of anatomical structures near the edges of the brain including the veins covering the cortical surface, spinal cord and nerve tracts near the mouse brain boundaries. The advantages of high-quality maps, no need for brain volume extraction, and high reconstruction speed demonstrate autoQSM's potential for future applications.

# Introduction

Quantitative susceptibility mapping (QSM) is a relatively new MRI technique that measures the spatial distribution of magnetic susceptibility within an object (Acosta-Cabronero et al., 2016; Bilgic et al., 2012; Haacke et al., 2015; Li et al., 2016; Liu et al., 2015a; Liu et al., 2015b; Schweser et al., 2013; Shmueli et al., 2009; Wang and Liu, 2015; Wharton and Bowtell, 2010). QSM computes the susceptibility from the phase signal of gradient-recalled echoes (GRE) and typically requires several processing steps. These steps involve phase unwrapping, tissue volume (e.g. brain) extraction, background phase removal and solving an inverse problem relating the tissue phase to the underlying susceptibility distribution. Phase unwrapping can easily be performed using path-based (Jenkinson, 2003) or Laplacian-based (Li et al., 2011; Schofield and Zhu, 2003) algorithms. The widely-used automatic Brain Extraction Tools (BET) are typically model-based (Smith, 2002) or learning-based (Iglesias et al., 2011). Removal of background fields may be performed using a number of algorithms, including projection onto dipole fields (Liu et al., 2011a), SHARP processing and its variants (Schweser et al., 2011; Wu et al., 2012) and HARPERELLA (Li et al., 2013). However, estimating the susceptibility map from a local tissue field map is more complex. To account for regions where the amplitude of dipole kernel is small and its inverse is undefined, some algorithms use threshold-based masking or dipole kernel modification (Schweser et al., 2013; Wharton et al., 2010). These algorithms are efficient and easy to implement; however, they contain severe streaking artifacts and bias susceptibility values due to the information loss through the masking process, and a compromise must be made between noise amplification and the reduction of streaking artifacts. Streaking in the focal areas of objects with large susceptibility values, e.g. blood vessels, may be reduced by estimating the missing data using iterative (Sun et al., 2016; Tang et al., 2013; Wei et al., 2015) or compressed sensing (Wu et al., 2012) algorithms. However, these iterative methods are considerably slower than direct inverse via thresholding, and care must be taken on the assumptions made when selecting spatial priors to avoid over-regularization and the reduction of image contrast (Liu et al., 2012b; Liu et al., 2011b; Wharton and Bowtell, 2010).

A new class of QSM algorithms that directly relates the GRE phase signal to the unknown susceptibility distribution has been proposed recently (Chatnuntawech et al., 2017; Liu et al., 2017; Sun et al., 2018). By performing background phase removal and dipole inversion in a single step, these algorithms prevent potential error propagation across successive operations. For example, to eliminate the background phase removal step, one study proposed a single-step QSM reconstruction technique which combined single-kernel spherical mean value (SMV) filtering with dipole inversion using the Laplacian operator (Chatnuntawech et al., 2017). Others used total generalized variation (TGV) regularization to develop a single-step QSM model (SS-TGV-QSM) that mitigated the artifacts observed in total variation (TV)-based reconstructions (Chatnuntawech et al., 2017; Langkammer et al., 2015). These Laplacian-based QSM methods implicitly eliminate the background field. However, the practical implementation of the Laplacian requires a tradeoff between robustness to error amplification and the integrity of the cortical brain tissue (Chatnuntawech et al., 2017; Langkammer et al., 2015). Furthermore, the necessary erosion of the brain mask may prevent visualization of structures at the brain boundary. Recently, total field inversion (TFI) and least square norm (LN-QSM) methods were proposed to directly perform dipole inversion on the total field (Liu et al., 2017; Sun et al., 2018). However, these methods still need a mask to aid QSM reconstruction. Moreover, automatically generating an optimal mask is challenging, especially near the brain boundary, where large air-tissue or tissue-bone susceptibility differences can cause substantial signal loss on the magnitude images used to define the mask.

Using a mask that is too big can include noisy phase information and lead to streaking artifacts, while a mask that is too small results in non-visualized brain. Erosion of the brain mask can especially prevent visualization of important structures at the brain boundaries, e.g. human brain cortical vessels, spinal cord and nerve tracts of mouse brain.

Deep neural networks have been applied to iterative methods for solving variety of inverse problems (Oktem, 2017; Qin et al., 2018). Over the last few years, deep learning methods have been shown to outperform previous state-of-the-art machine learning techniques in several fields, computer vision being one of the most prominent cases. Deep networks have been also applied to medical image reconstruction, e.g., PET, CT and MRI (Han et al., 2018; Leynes et al., 2018; Zhu et al., 2018). Recently, Yoon et al. trained a neural network to predict high-quality COSMOS (Calculation of Susceptibility through Multiple Orientation Sampling) (Liu et al., 2009) QSM from single-head-orientation data (Yoon et al., 2018). However, this trained model for COSMOS QSM does not describe magnetic susceptibility anisotropy. Another study proposed a deep convolutional network that utilizes real-world single-orientation phase to solve the inverse problem from simulated phase to magnetic susceptibility (Rasmussen et al., 2018).

In this study, we propose to train a deep neural network that reconstructs QSM directly from the field map while maintaining the contrast resulting from brain tissue's magnetic susceptibility anisotropy. The proposed method, referred to as autoQSM, is iteration-free, skipping skull stripping thus enabling efficient reconstruction. The model is trained on subjects with ages ranging from 11 to 82 years old, and the ground truth is the reconstruction by the established two-step STAR-QSM method (Wei et al., 2015). We investigate the capability of the trained neural network to directly reconstruct QSM from the measured magnetic field shift. We demonstrate the feasibility of autoQSM for fast and high-quality QSM reconstruction without skull stripping and show that it preserves more tissues at the brain boundaries, e.g. blood vessels and spinal cord. Moreover, we validate the network by generating QSM maps of the *in vivo* mouse brain which has dissimilar tissue contrast to the human brain training data, suggesting that the network generalized and was able to learn the underlying mathematical relationship between magnetic field shift and magnetic susceptibility. The advantages of high-quality maps, no need for brain volume extraction, high reconstruction speed and recovering more cortical blood vessels demonstrate autoQSM's potential for future applications.

## Methods

### MRI data acquisition and processing

A total of 209 healthy subjects with ages ranging from 11 to 82 years old were included for training. The subjects were scanned at the Brain Imaging and Analysis Center (BIAC) at Duke University using a 3T scanner (MR 750, GE Healthcare, Milwaukee, WI) equipped with an 8-channel head coil. Imaging was carried out with the approval of the institutional review board (IRB) and informed consent from the adult subjects or from the legal guardians of the teenage subjects. The 21 teenage (ages 11-20, 10M/11F) subjects were scanned using a 3D GRE sequence with field of view (FOV) = 22×22 $cm^2$, matrix size = 256×256, flip angle (FA) = 20°, TR = 41 ms, $TE_1$/spacing/$TE_8$ = 4/2.82/29.4 ms, spatial resolution = 0.86×0.86×2 $mm^3$, and reconstruction spatial resolution = 0.86×0.86×1 mm³, SENSE factor = 2, total imaging acquisition time = 5.7 min . The 188 adult (ages 21-82) subjects were scanned with the following parameters: FOV = 22×22 cm², matrix size = 256×256, FA = 20°, TR = 34.6 ms, $TE_1$/spacing/$TE_8$ = 5.468/3/26.5 ms, spatial resolution = 0.86×0.86×1 mm³. SENSE factor = 2, total imaging acquisition time = 9.7 min. Detailed information about the subjects at each age interval is shown in Fig. 1.

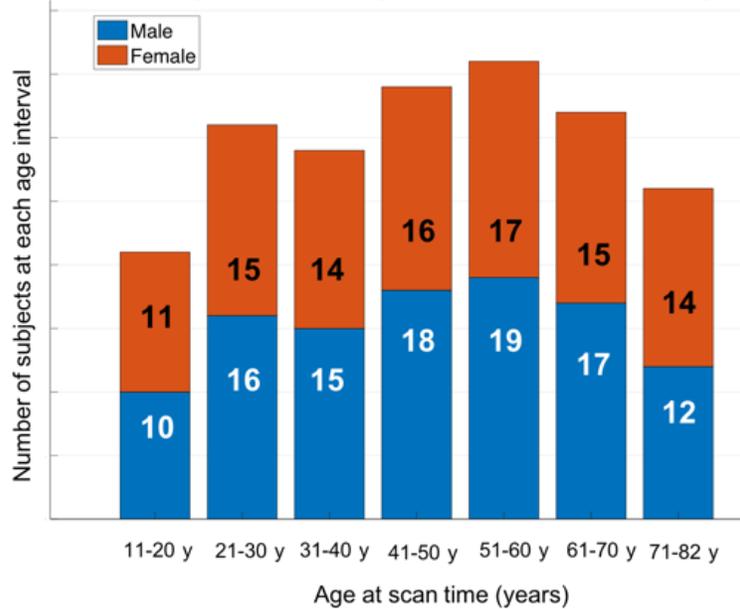

**Figure 1**. Number of subjects at each age internal for deep neural network training.

QSM reconstruction of the training dataset for the neural network was performed in STI Suite (https://people.eecs.berkeley.edu/~chunlei.liu/software.html). The sum of squares of GRE magnitude images across echo times ($\sum_{i=1}^{n} mag_i^2$), where n is the number of echoes, was used to mask and extract the brain tissue using the brain extraction tool (BET) in FSL (Smith et al., 2004). The raw phase was unwrapped using Laplacian-based phase unwrapping (Schofield and Zhu, 2003). The normalized total phase $\psi$ was calculated as: $\psi = \frac{\sum_{i=1}^{n} \omega_i}{\gamma \mu_0 H_0 \sum_{i=1}^{n} TE_i}$ where $\omega$ is the unwrapped phase. The normalized background phase was removed with the SMV method (Wu et al., 2012). The variable radius of the SMV filter increased from 1 pixel at the brain boundary to 25 towards the center of the brain with singular value decomposition truncated at 0.05 for the SMV filter during the deconvolution process (Wu, Li et al. 2012). Lastly, susceptibility maps were computed by inverting the filtered phase using the STAR-QSM algorithm (Wei et al., 2015; Wei et al., 2016).

**Deep network architecture**

Our network architecture is modified from an established architecture (U-net) (Ronneberger et al., 2015). The overall network architecture used in this study is summarized in Fig. 2. It consists of repetitive applications of 1) 3×3×3 convolutional layer, 2) batch normalization layer, 3) rectified linear unit (ReLU), 4) 2×2×2 convolution with stride 2, 5) 2×2×2 deconvolution with stride 2, 6) identity mapping layer that adds the left-side feature layer to the right side, and 7) a 1×1×1 convolution kernel as the last layer. The architecture can be divided into a contracting section and an expanding section. The left half of the architecture aims to compress the input path layer by layer, acting as an encoder, while the right half expands the path, acting as a decoder. The network has 15 convolutional layers in total and the largest feature size is 128. The network parameters including the depth of the layers were empirically optimized.

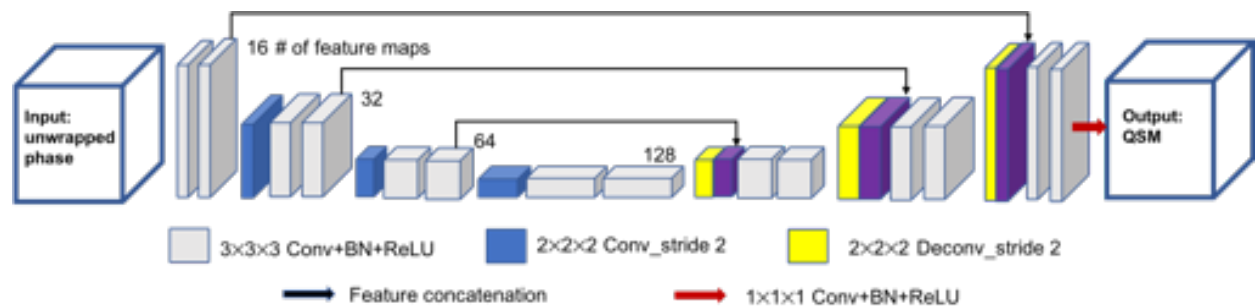

**Figure 2**. The schematic diagram of the neural network structure of autoQSM. A 3D U-net was implemented with 14 convolutional layers with kernel size of 3×3×3, 1 convolutional layer with kernel size of 1×1×1, 3 convolutional layers with kernel size of 2×2×2 applied with stride 2, 3 deconvolutional layers with kernel size of 2×2×2 applied with stride 2, and 3 feature concatenations.

The normalized 3D total phase images, $\psi$, were used as the input of the neural network and STAR-QSM images were used as the label. Out of the 209 healthy datasets, 42 subjects (6 subjects with 3F/3M from each age interval) were used as a validation set. Mean squared error (MSE) within the brain tissue between the reconstruction from the autoQSM and the label data served as the cost function for the optimizer, and it was minimized using the alternating direction method of multipliers (ADAM) optimizer (Eckstein and Fukushima, 1994). The learning rate decay was exponential with a factor of $10^{-4}$ every 600 steps until it reached $10^{-7}$. The batch size was set to 8 and the network converged after 100 epochs. To avoid overfitting, dropout was used to randomly turn off neurons with a rate of 10% (Srivastava et al., 2014). The proposed network structure was implemented using Python 3.6.2 and Tensorflow v1.4.1 using NVIDIA 1080TI GPU. The total training time was approximately 24 hours.

**Evaluation of autoQSM**

To test the network's ability to reconstruct QSM directly from total phase images, different datasets acquired at different sites were used as test datasets. These datasets had different acquisition parameters from the training data, so the following experiments would test autoQSM's ability to learn the underlying physical principle of the QSM reconstruction rather than simply the anatomy.

**Experiment 1**: Numerical brain phantom was built according to a previous QSM study (Chatnuntawech et al., 2017). The phantom was generated with the following susceptibility values (SI units): hippocampus, 0.05 ppm; hypothalamus, 0.05 ppm; medulla oblongata, 0.05 ppm; white matter, –0.03 ppm; cerebellum, –0.0065 ppm; pons, –0.0065 ppm; thalamus, –0.0065 ppm; midbrain, –0.0065 ppm; cerebrospinal fluid (CSF), 0 ppm; skull, –2.1 ppm. The magnetic susceptibility sources at 0.6 ppm were included to resemble subcutaneous fat without the chemical shift effect. The magnetic susceptibility sources at 9.2 ppm were included within the head to mimic internal air in the nasal cavity. The skull, fat and air act as the background susceptibility sources compared to brain tissues. The resulting phantom was convolved with the dipole kernel to generate the total phase. The total phase defined within the brain mask was as the input of the trained neural network and the result was compared to those computed by the truncated k-space division (TKD) (Wharton and Bowtell, 2010) and the improved sparse linear equation and least-squares (iLSQR)-algorithm (Li et al., 2015).

**Experiment 2**: Six subjects were acquired using a 3D GRE sequence on a GE 3T Hdxt scanner equipped with an 8-channel head coil with the following parameters: FOV = 25.6×25.6 cm$^2$, matrix size = 256×256, FA = 20°, TR = 41 ms, TE$_1$/spacing/TE$_{16}$ = 3.2/2.2/36.2 ms, spatial resolution = 1×1×1 mm$^3$. The same scans were repeated at three additional head orientations with respect to the B$_0$ field, SENSE factor = 2, total imaging acquisition time = 12 min per orientation. The four orientations were used to perform COSMOS QSM reconstruction. In addition, TKD and iLSQR QSM images were generated from the filtered phase for comparison. For the TKD method, the threshold equals to 0.2 as suggested in the literature (Shmueli et al., 2009). For the iLSQR method, the parameters were set as suggested in the original paper (Li et al., 2015). To assess its performance, the results of autoQSM were compared with STAR-QSM, TKD, iLSQR and the gold-standard COSMOS QSM. We used several quantitative metrics assessed by the 2016 QSM Reconstruction Challenge to evaluate the reconstruction quality of these QSM algorithms. The metrics were normalized root-mean-squared error (RMSE), high-frequency error norm (HFEN), and structure similarity index (SSIM) (Langkammer et al., 2018). Note that all the metrics were assessed within the brain tissue mask. To further quantify the accuracy and consistency of the QSM maps, region-of-interest (ROI) analysis was performed. ROIs were extracted by registering a QSM atlas (Zhang et al., 2018) to the reconstructed QSM images. The QSM dataset was treated as the target brain and the QSM atlas was registered using large deformation diffeomorphic mapping (LDDMM) (Beg et al., 2005). Deep gray matter (DGM) ROIs, including caudate nucleus (CN), putamen (PUT), globus pallidus (GP), red nucleus (RN), substantia nigra (SN), and representative white matter (WM) ROIs, including internal capsule (IC), corpus callosum (CC), optic radiation (OR), were defined using the QSM atlas. The mean and standard deviation for each ROI was calculated for the QSM maps reconstructed by different methods.

**Experiment 3**: AutoQSM was also tested using the data provided by the 2016 QSM Reconstruction Challenge (Langkammer et al., 2018). RMSE, HFSN, SSIM, and ROI error were calculated with respect to the susceptibility tensor component $\chi_{33}$.

**Experiment 4**: This experiment aimed to test autoQSM's performance on infant and child brain datasets, which exhibit lower iron deposition and less myelination compared to the adult brain thus significantly different contrast. Imaging was carried out with approval of IRB and parental consent for babies and children. Ten 2-years-old infant subjects were scanned with the following parameters: FOV = 220×220 mm$^2$, matrix size = 220×220, TR = 50 ms,

TE$_1$/spacing/TE$_{16}$ = 2.9/2.9/46.4 ms, and spatial resolution = 1×1×1 mm$^3$, SENSE factor = 2, total imaging acquisition time = 11 min. Infant earmuffs were used for hearing protection, and possible motion artifacts were mitigated by immobilization with a cotton pillow. An experienced neonatologist and a neuroradiologist were in attendance throughout the imaging process. Fifteen children (ages 5-7) were scanned with the following parameters: FOV = 220×220 mm$^2$, matrix size = 384×384, TR = 38 ms, TE$_1$/spacing/TE$_7$ = 4.35/4.76/32.91 ms, and spatial resolution = 0.58×0.58×1 mm$^3$, SENSE factor = 2, total imaging acquisition time = 12.3 min.

**Experiment 5**: The fifth experiment attempted to explore the clinical applicability of autoQSM to data from patients with brain lesions, which were not present in the subjects seen by the network during training. Fifteen multiple sclerosis (MS) patients were scanned with the following parameters: FOV = 220×220 mm$^2$, matrix size = 256×256, TR = 54 ms, TE$_1$/spacing/TE$_8$ = 3/4.18/32.3 ms, and spatial resolution = 0.86×0.86×1 mm$^3$, SENSE factor = 2, total imaging acquisition time = 13.5 min. Fifteen patients with brain hemorrhage was scanned with the following parameters: FOV = 220×220 mm$^2$, matrix size = 256×256, TR = 43 ms, TE$_1$/spacing/TE$_8$ = 3.16/4.85/37.1 ms, and spatial resolution = 0.86×0.86×1 mm$^3$, SENSE factor = 2, total imaging acquisition time = 11.9 min

**Experiment 6**: We also applied the trained network to total phase maps of fifteen *in vivo* mouse brain data that were scanned using a 7T 20-cm-bore magnet (Bruker BioSpec 70/20 USR, Billerica, MA, USA) interfaced to an Avance III system. A high-sensitivity cryogenic radiofrequency coil was used for transmission and reception (Bruker CryoProbe). The mice were scanned using a 3D spoiled-gradient-recalled (SPGR) sequence with the following scan parameters: TR = 250 ms, TE$_1$/ΔTE/TE$_{10}$ = 3.72/5.52/53.4 ms, FA = 35°, FOV = 19.2×14.4×9.6 mm$^3$ with 87 μm isotropic resolution, total imaging acquisition time = 90 min. Data acquisition was respiratory gated with two pulse sequence repetitions per respiratory cycle.

**Experiment 7**: Three subjects were scanned using a 3D fast low angle shot (FLASH) sequence on a Siemens Magnetom Terra scanner equipped with an 32-channel head coil with the following parameters: sagittal view, FOV = 19.2×19.2 cm$^2$, matrix size = 320×320, FA = 10°, TR = 30 ms, TE$_1$/spacing/TE$_4$ = 2.3/6.9/23 ms, spatial resolution = 0.6×0.6×0.6 mm$^3$, GRAPPA factor = 2, total scan time = 25.6 min. To assess autoQSM's performance on image the magnetic susceptibility of spinal cord, the results of autoQSM were compared with STAR-QSM's results.

**Results**

Fig. 3 shows the testing results on numerical phantom susceptibility model using different methods. The difference map between the results of autoQSM and true susceptibility shows negligible susceptibility differences related to brain tissues. Compared to TKD and iLSQR, autoQSM delivers substantially lower error level with RMSE of 58%, 76% and 83% respectively.

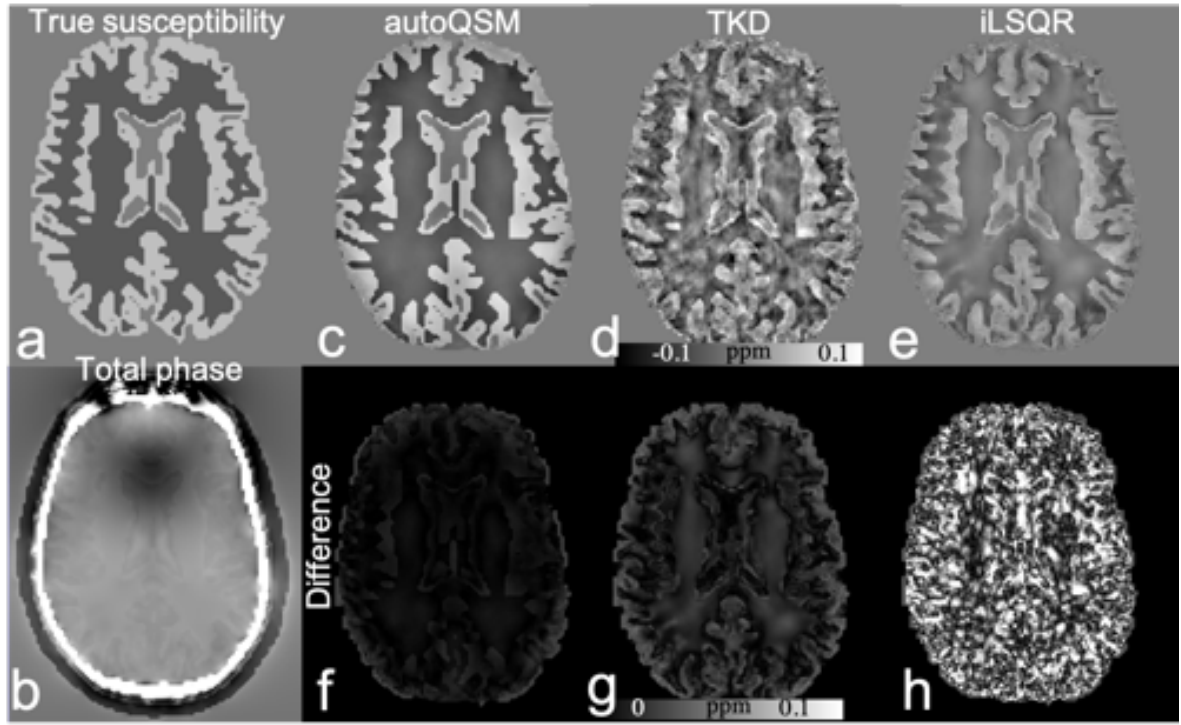

**Figure 3**. Results of autoQSM, TKD and iLSQR for QSM reconstruction on the brain numerical phantom susceptibility model as well as the difference with respect to the ground truth. The input is the masked total phase map from a simulated background field superimposed onto the simulated brain phantom. Compared to TKD and iLSQR, autoQSM delivers substantially lower error level.

Fig. 4 shows the three orthogonal views of raw phase, total phase and QSM images on one representative healthy subject using the five methods. It is clear from Fig. 4 that autoQSM can effectively recover the cortical tissues, such as vessels. While TKD-QSM and iLSQR showed a substantially noisy susceptibility contrast between the cortical gray and white matter. AutoQSM preserves better susceptibility delineation between gray and matters similar to COSMOS, benefiting the improved image quality via regression of deep learning. The quantitative metrics, e.g., RMSE, HFEN, and SSIM of the five reconstruction methods are summarized in Table 1. AutoQSM results achieved the lowest NMSE, lowest HFEN and highest SSIM, suggesting better performances based on these criteria than the TKD-QSM and iLSQR methods.

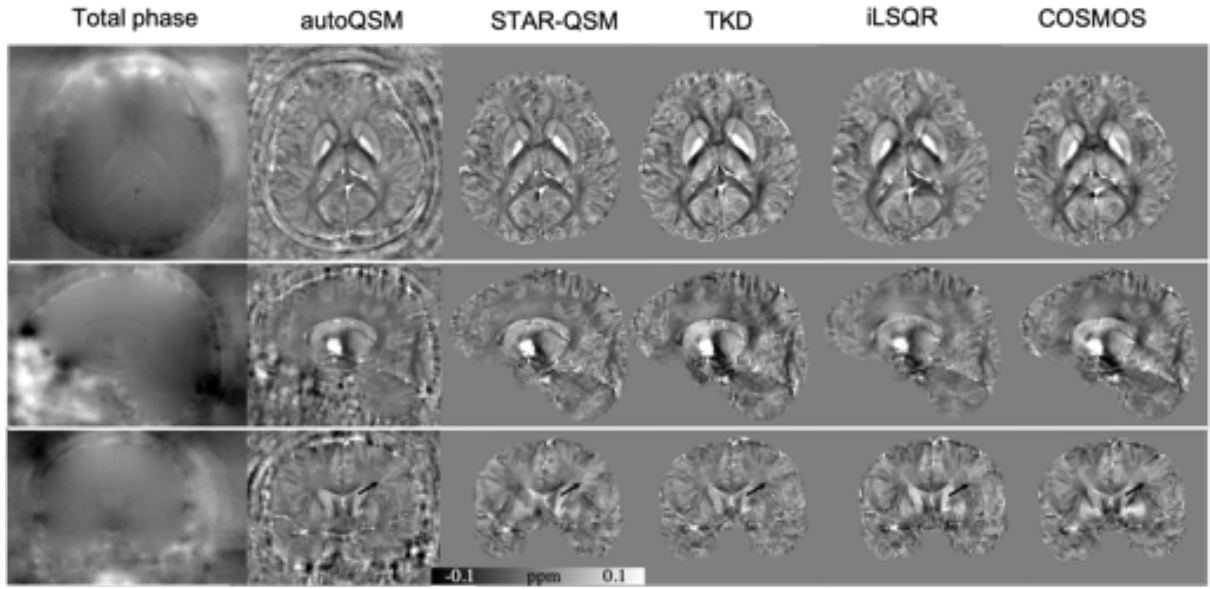

**Figure 4**. Comparison of different QSM reconstruction methods on a healthy subject referenced to COSMOS QSM. Arrows pointed to the cortical gray and white matter delineation can be visible on autoQSM's results which hardly seen on TKD and iLSQR QSM images.

**Table 1**. Quantitative performance metric, RMSE, HFEN, and SSIM from the four different QSM reconstruction methods referenced to COSMOS QSM. AutoQSM shows better performances in all criteria than other QSM methods.

|  | RMSE (%) | HFEN(%) | SSIM |
|---|---|---|---|
| TKD | 75.6 | 75.1 | 0.88 |
| iLSQR | 74.5 | 73.3 | 0.86 |
| STAR-QSM | 72.6 | 68.8 | 0.91 |
| autoQSM | 72.2 | 68.8 | 0.91 |

With respect to the mean susceptibility and standard deviation of the representative ROIs as shown in Fig. 5a, the autoQSM results show comparable values when compared to STAR-QSM and the gold-standard results of COSMOS. The autoQSM results also show better accuracy than the two other QSM methods evaluated.

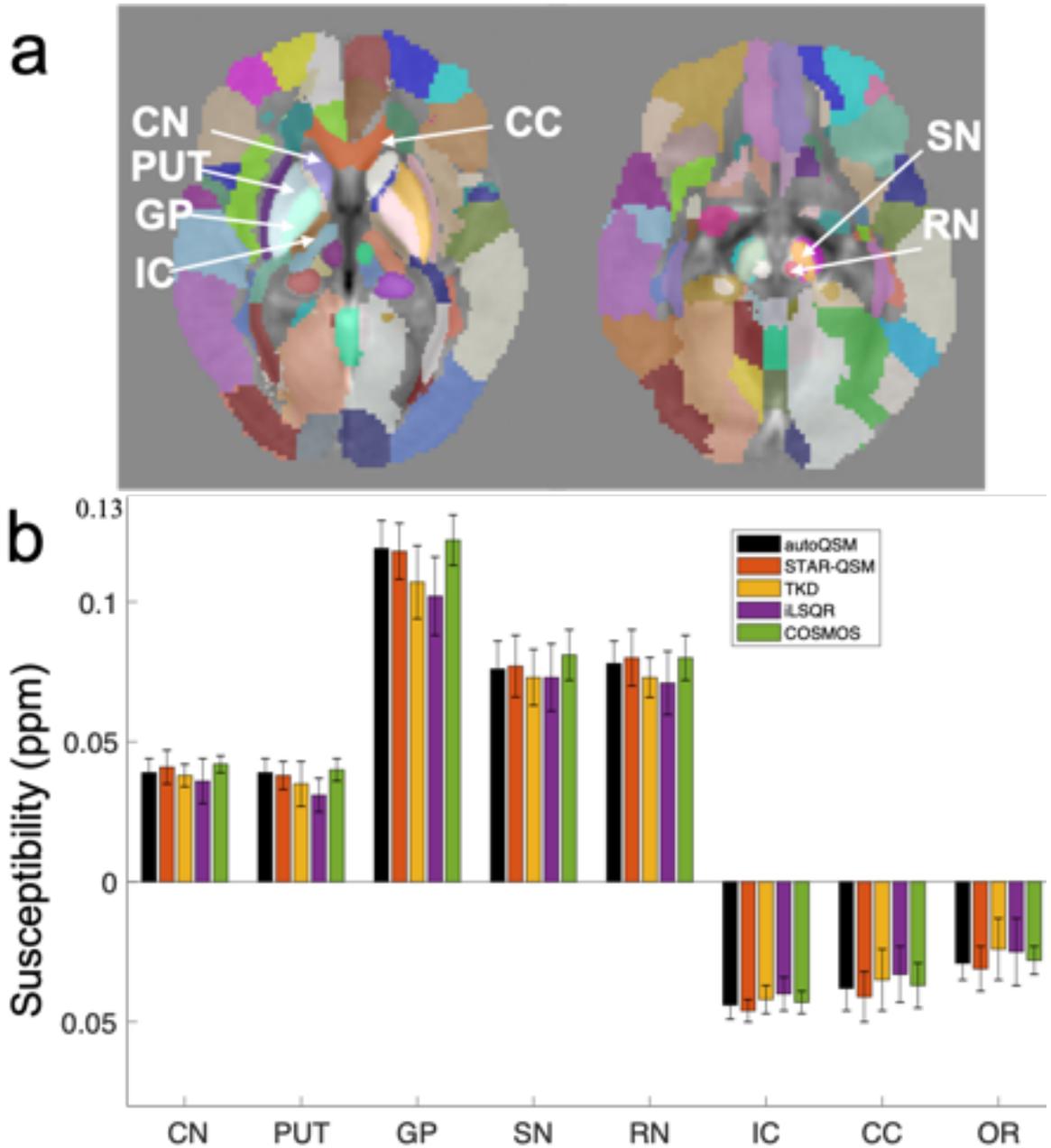

**Figure 5**. ROI analysis of the five different methods. The susceptibility values of the ROIs (CN, PUT, GP, SN, RN, IC, CC, OR) are plotted, the autoQSM's results match well with the gold-standard COSMOS QSM results. Data are presented as mean ± standard deviation.

Using data provided from the 2016 Challenge and $\chi_{33}$ as the ground truth, we compared the performance metrics between STAR-QSM and autoQSM. AutoQSM returned comparable scores to STAR-QSM for RMSE and HFEN metrics, and ROI-level errors equivalent to those observed for STAR-QSM and $\chi_{33}$. Qualitatively, we note that autoQSM yielded susceptibility values and contrast at brain edges that are lost in STAR-QSM and $\chi_{33}$, as shown in Fig. 6d. Note that the checkerboard and ringing artifacts near the cortical surface were observed on the autoQSM's results. These artifacts are due to noisy unwrapped phase values with checkerboard pattern outside of the brain that acquired using the simultaneous multiple slice (SMS) sequence. However, these artifacts are not present in the data acquired using conventional GRE sequences. The comparison of unwrapped phase images acquired using SMS and GRE sequences are shown in Fig. S1 in the supplemental material.

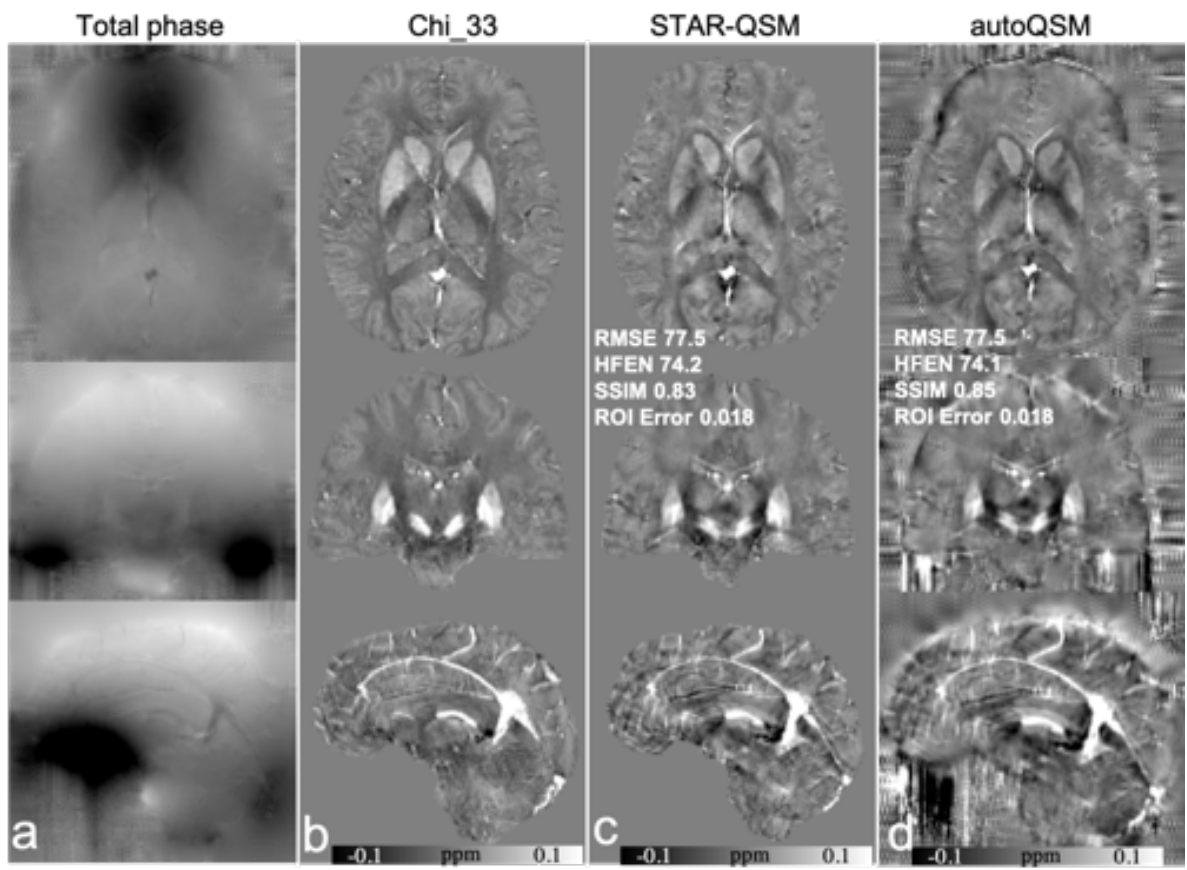

**Figure 6**. Comparison of Chi_33, STAR-QSM and autoQSM using the 2016 QSM Challenge data. (a) total phase, (b) Chi_33, (c) STAR-QSM, (d) autoQSM. Note that the noisy phase values with checkerboard pattern outside of the brain in d are due to artifacts on the unwrapped phase images, rather than autoQSM reconstruction. The RMSE, HFEN, SSIM are calculated within the mask defined based on Chi_33.

The autoQSM method was applied to the infant, child and adult subjects which were not included in the training dataset. The predicted results by autoQSM revealed comparable contrasts to those of STAR-QSM. As shown in the difference maps (Fig. 7c), there are negligible susceptibility differences related to gray and white matter. The clear differences at the edge of the brain were caused by blood vessels that were predicted by the trained neural network but lost in STAR-QSM. Similar results were observed when autoQSM was applied to the patients with MS lesions and hemorrhage. Another dataset was shown in Fig. S2 in the supplemental material.

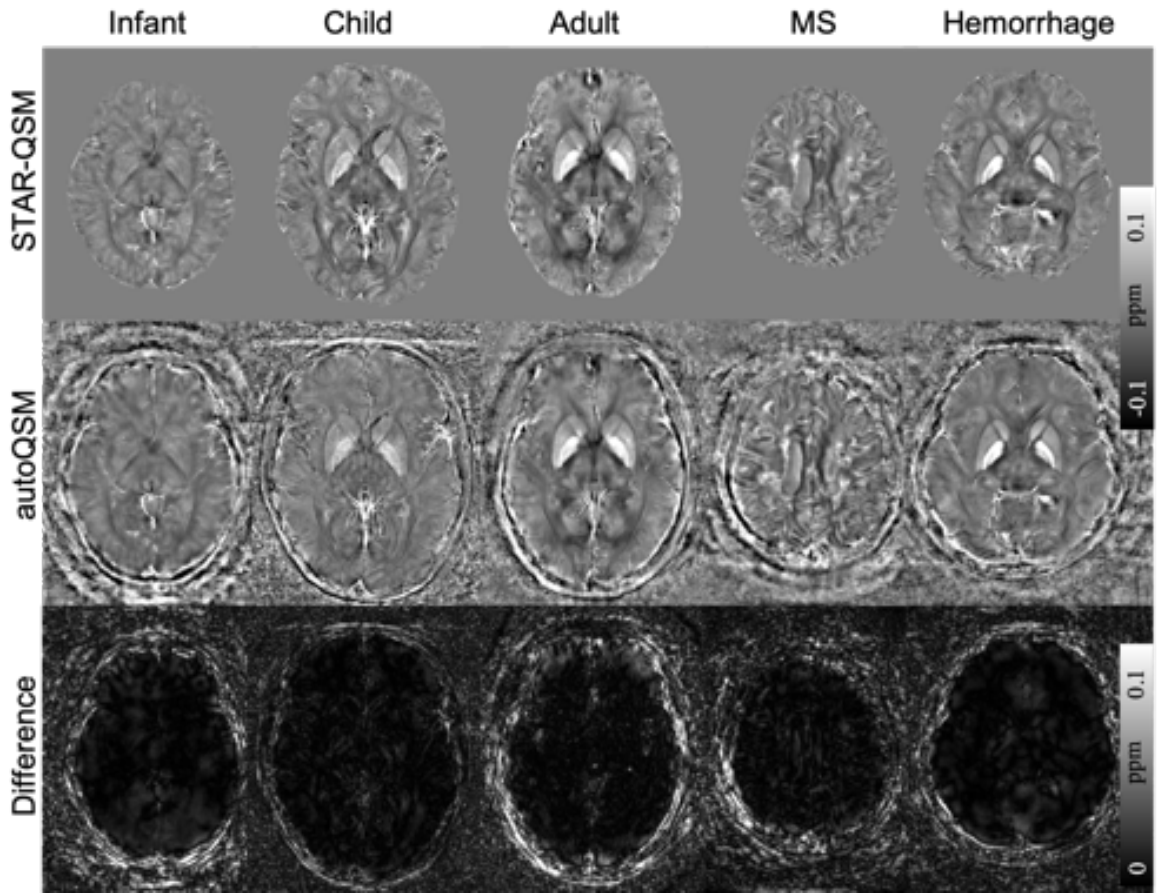

**Figure 7**. Comparison of QSM images computed using STAR-QSM and autoQSM methods. The last row shows the difference maps between QSM images reconstructed by the two methods.

It is well known that brain volume extraction from *in vivo* mouse brain MRI images is more complex because the brain is surrounded by tissues that have similar image intensity. In addition, the gap between the brain and non-brain tissue is very narrow. At some locations, the edges of the brain cannot even be identified at an isotropic spatial resolution of 86 μm, as shown in Fig. 8a. Consequently, human brain extraction techniques are error prone when applied to mouse brain MRI. For example, the medulla region including nerve tracts as shown in Fig. 8d is significantly eroded. Fig. 8c illustrates autoQSM's remarkable capability of preserving the cortical regions of the *in vivo* mouse brain without skull striping during reconstruction. Red arrows pointed to cortical regions with shadowing artifacts that are significantly reduced using the trained network. Yellow arrows point to brain erosion that can be recovered using autoQSM. For example, we can observe the white matter tracts as pointed by black arrows in Fig. 8c, which is completely inaccessible on the STAR-QSM images.

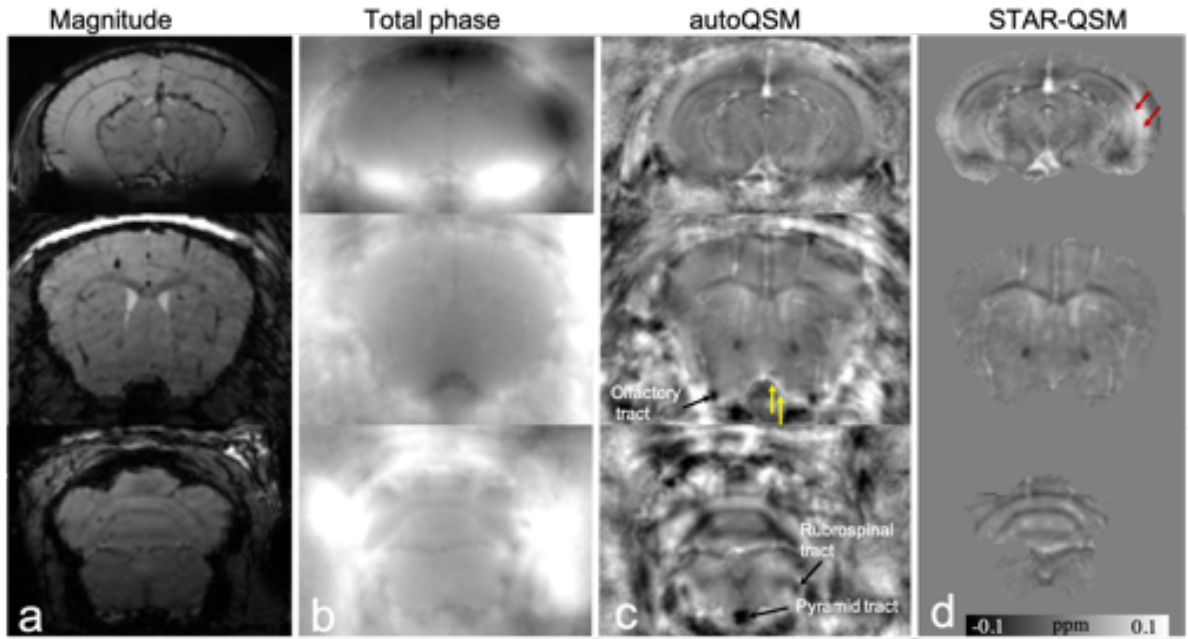

**Figure 8**. Representative axial slices of QSM images computed using STAR-QSM and autoQSM methods in an *in vivo* mouse brain. (a) Magnitude images, (b) total phase maps, (c) QSM images predicted using trained neural network, (d) QSM image reconstructed using STAR-QSM. Red arrows point to reduced artifacts by the trained neural network. Yellow arrows point to paramagnetic susceptibility of blood vessel preserved near the cortex. Black arrows point to the recovered nerve tracts near the edges of the brain revealed by autoQSM while eroded by skull stripping.

Additionally, the trained neural network may have the potential to imaging the magnetic susceptibility of tissues near the neck, e.g. spinal cord. AutoQSM's images show clear susceptibility contrast between gray and white matter. Black arrows pointed to the diamagnetic lateral white matter tracts while white arrows pointed to the gray matter regions which have relatively paramagnetic susceptibility values. All three healthy subjects show the consistent contrast between white matter and gray matters in Fig. 9. However, these spinal cord regions are significantly eroded during skull stripping procedure as shown by STAR-QSM images. We further manually added the eroded mask regions back to the brain mask generated using FSL BET. We computed STAR-QSM's results using the mask covering the spinal cord. The comparison of magnetic susceptibility of spinal cord using two processing is shown in Fig. S3. Similar susceptibility contrast was observed in the spinal gray and white matters between Fig. S3b and Fig. S3c except that the erosion exists in STAR-QSM due to background phase removal procedure, confirming the reconstruction accuracy of autoQSM for imaging white and gray matters of spinal cord.

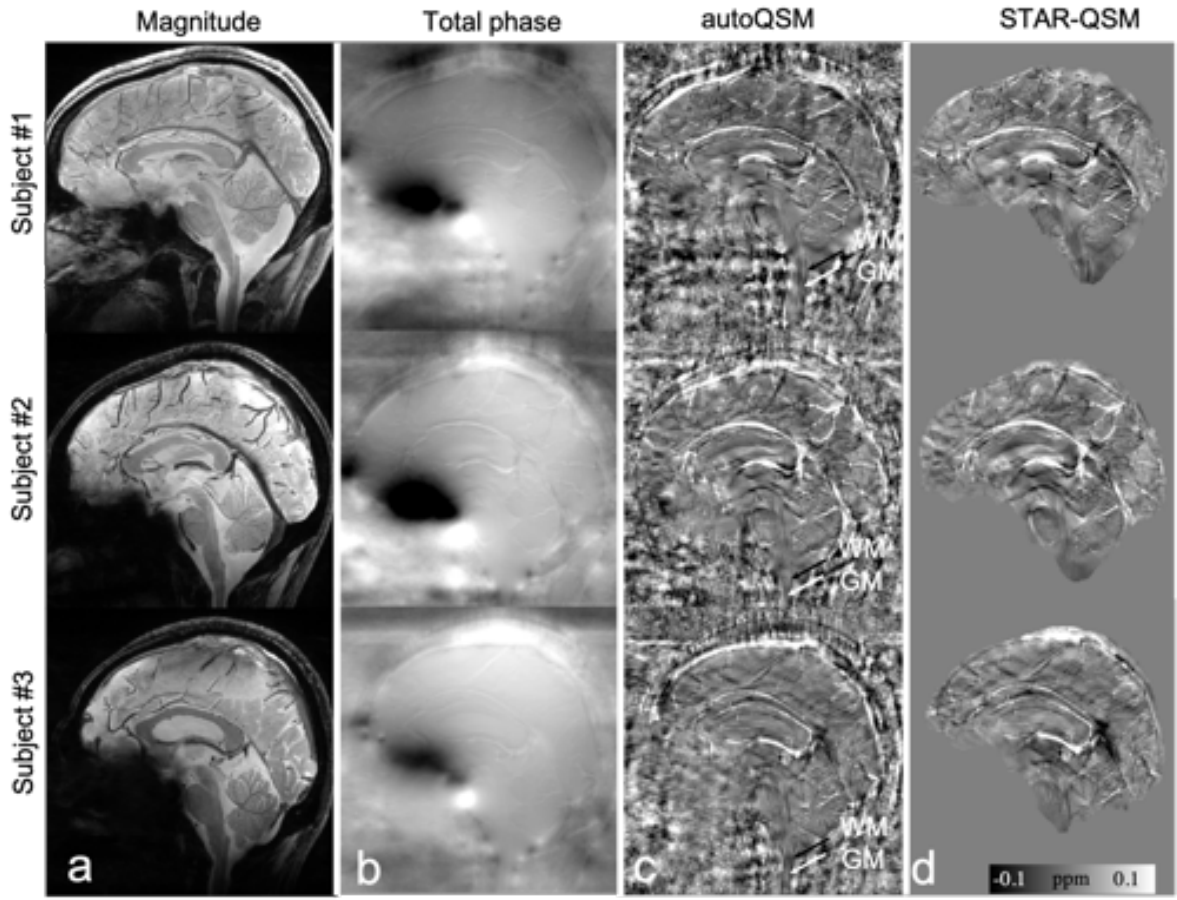

**Figure 9**. Comparison of QSM reconstructed by STAR-QSM and autoQSM on three healthy volunteers scanned at sagittal view with spinal cord included at 7 T. Black and white arrows point to white matter and gray matter in spinal cord respectively. Note that the paramagnetic susceptibility of gray matters is surrounded by diamagnetic white matters in the spinal cord.

One main advantage of the autoQSM is the fast reconstruction speed. The average reconstruction time was only 5 ± 0.8 s measured in a GPU, which was much fast than STAR-QSM (146 ± 16.2 s; with background removal measured in a CPU) and iLSQR (360 ± 32.5 s; with background removal measured in a CPU).

**Discussion**

In this study, we constructed a deep neural network that performs QSM reconstruction from total phase images without brain volume extraction. One significant advantage is the high computational efficiency of the trained neural network achieved by combining two techniques: (i) eliminating the need of brain skull stripping during QSM reconstruction, and (ii) end-to-end QSM processing computed by GPU. Compared to conventional QSM reconstruction methods involving background phase removal followed by dipole field inversion, our results show better quality of reconstructed susceptibility maps. Especially, skipping the skull stripping for QSM reconstruction significantly improves the robustness of QSM reconstruction to brain volume extraction bias. In addition, the training data used in this study covers a wide age range (11 to 82 years old), which is important for high reproducibility in longitudinal studies. The preliminary results tested on infant brain data, *in vivo* mouse brain data, and patients with brain lesions suggest that autoQSM can be applied to brain data dissimilar to the training data, which suggests that the network has generalized the underlying principles of QSM inversion. Finally, the high computational efficiency allows for clinical routine QSM reconstruction within a few seconds.

Previously, deep neural networks have shown the ability to reconstruct QSM from simulated phase, similarly to any other dipole deconvolution methods (Rasmussen et al., 2018). In this study, we reconstructed a neural network to incorporate background phase removal and compared the susceptibility reconstructions from autoQSM to the state-of-the-art QSM algorithms. Other single-step QSM reconstruction method, e.g. the SS-TGV-QSM method combined Laplacian-based background phase removal and local field inversion into a single step. However, Laplacian-based methods suffer from brain erosion since it is implemented using the finite difference operator or the spherical kernel operator, both require the ROI mask to be eroded. The LN-QSM and TFI methods perform dipole inversion directly on the total field instead of on the filtered phase and thus avoid the Laplacian operator, but still require brain masks to aid QSM reconstruction. Additionally, it was shown that the reconstruction speed and the quantification accuracy are both influenced by the choice of the preconditioner in TFI and regularization parameters in LN-QSM (Liu et al., 2017; Sun et al., 2018). In this study, the trained neural network enables end-to-end single-step QSM processing and it does not require explicit regularization parameters.

Brain extraction is required for all the existing QSM reconstruction methods. A number of automated brain extraction algorithms have been developed using morphology, morphology combined with edge detection (Shattuck and Leahy, 2001), deformable models (Smith, 2002), graph cuts, watershed and others. Each algorithm has its merits and pitfalls. More generally, the accuracy of brain extraction depends on the segmentation algorithms and parameters (Iglesias et al., 2011; Smith, 2002). Each method either keeps some non-brain tissue or removes extra brain tissue. Thus, using brain extraction is problematic when applying QSM to a large cohort with varying scanning parameters and manual adjustment of program parameters and manual editing of extraction results are inevitable. Brain extraction is especially problematic for baby brain MRI images because they lack edge information and contain ambiguous tissue information. Similarly, the *in vivo* mouse brain has a narrow gap between skull and brain tissue, which hampers the ability of QSM to evaluate the magnetic susceptibility of brain tissue. With autoQSM, brain QSM maps without brain extraction is possible by direct prediction on the total phase within the whole FOV. Furthermore, superficial veins can be recovered by the trained neural network. The recovered magnetic susceptibility at the cortical surface may extend potential QSM applications to functional QSM imaging (Sun et al., 2017) and quantitative full brain susceptibility venography (Buch et al., 2019). In addition, the recovered magnetic susceptibility of spinal cord may extend potential QSM applications to investigate the magnetic susceptibility between spinal cord lesions and brain lesions for Multiple Sclerosis patients.

Recently, there was a proposal of using deep neural network (QSMnet) to predict high quality COSMOS QSM maps from filtered phase acquired from a single orientation (Yoon et al., 2018). This result overcomes the drawback of long scanning time for the multiple head orientation acquisitions. Although the QSM maps produced by COSMOS have high quality with higher SNR, the disadvantage of the COSMOS model is that it does not account for anisotropy of magnetic susceptibility and structural tissue anisotropy. The comparison of QSM images reconstructed using QSMnet and autoQSM is shown in Fig. S4 in the supplemental material. Quantification of susceptibility anisotropy in white matter is crucial for investigation of myelin membrane lipids (Li et al., 2017; Li and van Zijl, 2014; Li et al., 2012; Liu, 2010; Liu et al., 2012a). It is has been reported that prenatal alcohol exposure significantly reduces susceptibility anisotropy of the white matter (Cao et al., 2014). Another study proposed using deep neural networks trained to solve the inverse problem from simulated phase to magnetic susceptibility (Rasmussen et al., 2018). The input data for training was created by convoluting the labeled synthetic real-world data with the dipole kernel following well-posed forward solutions. The trained network was then used to solve an ill-posed field-to-susceptibility inversion. However, the quantification accuracy of the trained model using simulated data needs further investigation.

The measured magnetic susceptibility of cortical surface blood depends on flow velocity, oxygenation level of hemoglobin, and the angle between the vessel and $B_0$ field. As shown in the current study, the magnetic susceptibility of blood vessels near the edges of the brain were recovered by autoQSM. From the difference map (Fig. S5), there were negligible susceptibility differences related to brain tissues between autoQSM and STAR-QSM. In contrast, clear differences near the brain boundary by the blood vessels do appear in the difference maps. These images show that autoQSM produces similar susceptibility values of the brain tissue to STAR-QSM while preserving the cortical vessels, suggesting autoQSM's potential to recover the cortical vessels. In the future, quantitative full brain susceptibility venography needs to be investigated.

The test datasets currently used in this study have different spatial resolutions, e.g. 1 mm isotropic spatial resolution in Experiment 1; 0.86×0.86×1 $mm^3$ used in Experiment 4, 87 μm isotropic resolution used in experiment 5, 0.6 mm isotropic resolution in Experement 7 and 0.86×0.86×2 $mm^3$ in Experiment 8 and with different matrix sizes. Different spatial resolutions will alter the SNR level of phase images which may bring some error in the background. The matrix size does not have any effects on the predicted images since the neural network was trained patch by patch. We may note to users that the test dataset should not have a large slice thickness size (e.g., larger than 4 mm) since the training and prediction patch is 64×64. Any matrix dimension smaller than 64 will corrupt the currently trained model. Also, the image orientation should be consistent between training and test data. In addition, the input unwrapped phase images should be normalized both for training and testing.

We observe the residual susceptibility contrast left outside of the brain. The autoQSM maps are not intended to estimated susceptibility of the air outside the head or regions without data support, except for special cases such as small air pockets surrounded by tissues. While phase unwrapping methods extrapolated phase values outside of the brain tissue resulting in the observed susceptibility contrast outside of the brain predicted by the neural network. The phase pattern and signal-to-noise ratio (SNR) of the unwrapped phase images also vary at different acquisition sites, depending on the phase reconstruction methods (e.g., coil combination method), severity of motion artifacts and phase unwrapping algorithms. As shown in Fig. S6, the phase pattern has huge difference between different MR sites. The neural network cannot distinguish the brain and non-brain regions from the unwrapped phase image. Thus, the predicted QSM image keep background contrasts when the background has non-zero unwrapped phase values. The mask can be applied afterwards to mask out the non-brain tissues only for visualization purpose, as shown in Fig.S7 in the supplemental material

The trained deep neural network performs QSM reconstruction from total phase images which functions as the background phase removal method. The autoQSM's performance for background phase removal on the numerical brain phantom experiment was shown in Fig.S8 in the supplemental material.

**Limitations**

The trained neural network has some limitations. One of the biggest challenges of deep networks is that they are difficult to characterize conceptually. The design of the autoQSM architecture and the training parameters require empirical tuning of the network structure.

The trained neural network was applied to the unwrapped phase image, assuming the phase wraps were accurately removed. Further research is necessary to fully explore the capability of deep neural network to reconstruct QSM from raw phase with phase wraps. However, the location and number of phase wraps are highly dependent on the scan parameters. The phase wrap pattern varies significantly if the brain tilts at different angles with respect to $B_0$ field. All these factors challenge the performance of deep neural network to reconstruct QSM from raw phase images. Alternatively, we can extend the training dataset to a much larger scale including datasets acquired at different echo times, different head rotations, and at different field strengths. We expect that more research will explore this possibility for QSM reconstruction in the future.

## Conclusion

Our results demonstrate a powerful new paradigm for QSM reconstruction without the need for brain volume extraction, which is implemented with a deep neural network that learns the underlying mathematical relationship between total field and the magnetic susceptibility. Quantitative and qualitative comparisons demonstrate that autoQSM has superior image quality compared to other QSM methods. Additionally, the autoQSM maps show its potential to explore the magnetic susceptibility of whole brain vasculature, spinal cord and cortical nerve tacts of mouse brain. The advantages of high-quality maps, no need for brain volume extraction, and high reconstruction speed demonstrate autoQSM's potential for future applications.

## Acknowledgements

This study is supported in part by the National Institutes of Health through grants NIMH R01MH096979 and U01EB025162. We thank Woojin Jung and Prof. Jongho Lee for QSMnet reconstruction used in the Figure S4 in the supplemental material.

# Supplemental materials

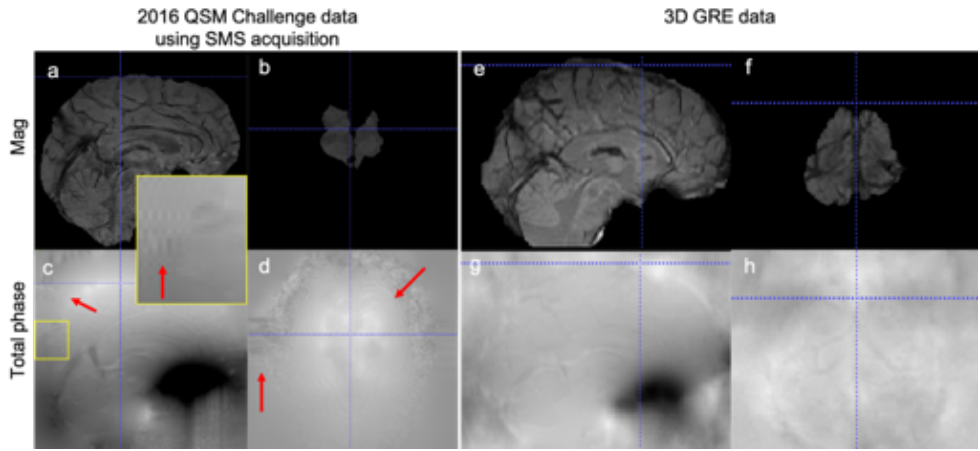

Fig. S1. Comparison of the unwrapped phase images acquired using simultaneous multiple slice sequences and conventional gradient echo sequences.

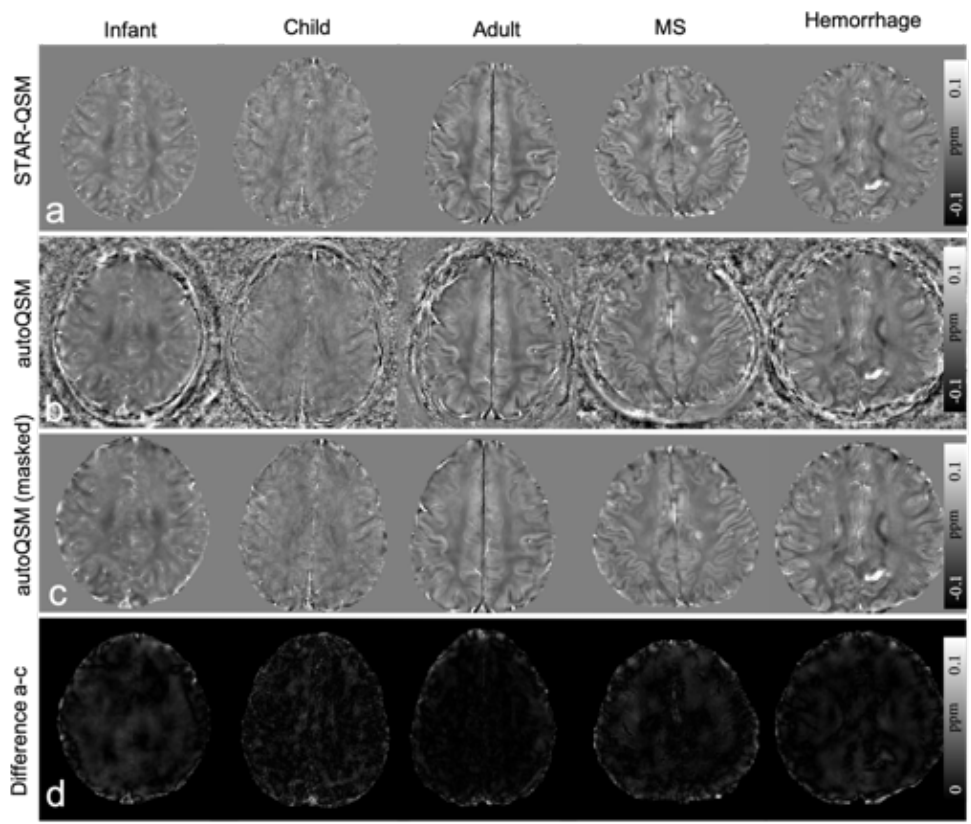

Fig. S2. Representative susceptibility images calculated using STAR-QSM and autoQSM are demonstrated on different datasets. C are the masked autoQSM Images by applying the derived mask from STAR-QSM images to b. d are the difference images between a and c.

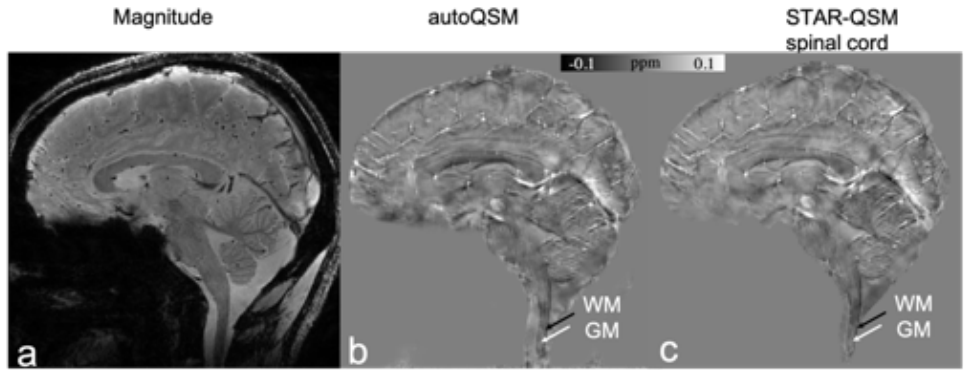

Fig. S3. Sagittal views of the QSM images showing the magnetic susceptibility of spinal cord. The predicted QSM images using autoQSM show strong contrast between gray and white matters. C is the results calculated using STAR-QSM by manually adding the eroded mask regions back to the brain mask generated using FSL BET.

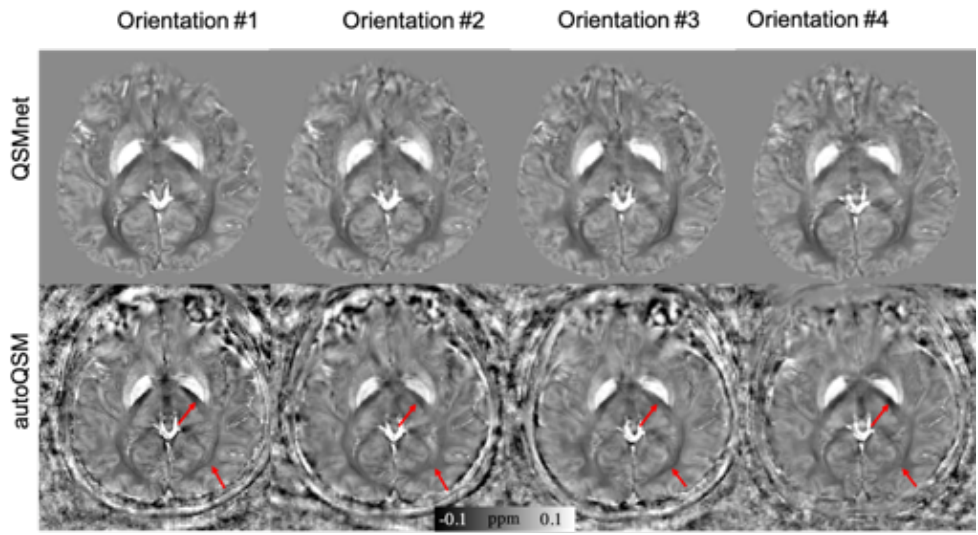

Fig. S4. Axial views of the QSM images from four head orientations. Compared to the QSMnet's results, the predicted QSM images using autoQSM show strong orientation dependency of magnetic susceptibility within the white matter. The QSMnet's results show a more consistent susceptibility contrast between gray-white matters across the orientations (Yoon et al., Neuroimage, 179(1):199-206, 2018).

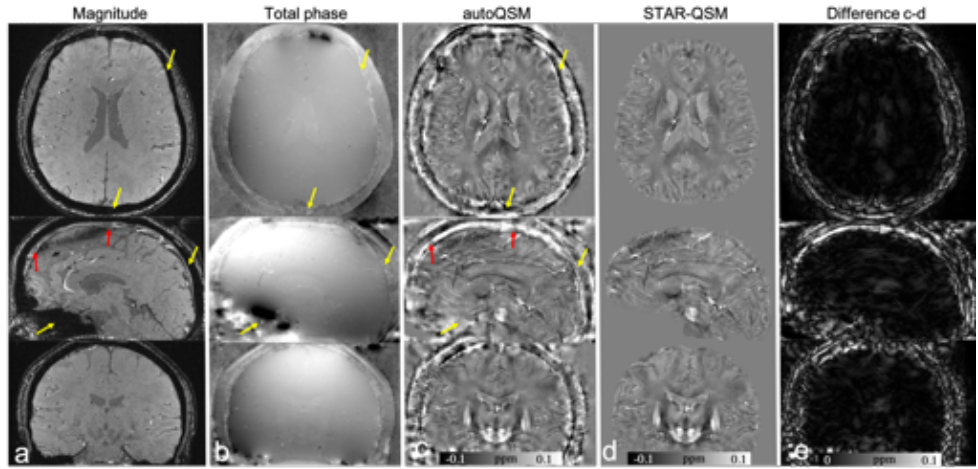

Fig. S5. Results of autoQSM for imaging the cortical vessels on a high-resolution dataset. (a) magnitude images; (b) total phase images; (c) autoQSM's results; (d) STAR-QSM's results; (e) difference maps between c and d. Red arrows point to diamagnetic susceptibility of blood vessel and yellow arrows point to inaccurate phase measurements of skull and air resulting severe dark and bright susceptibility artifacts.

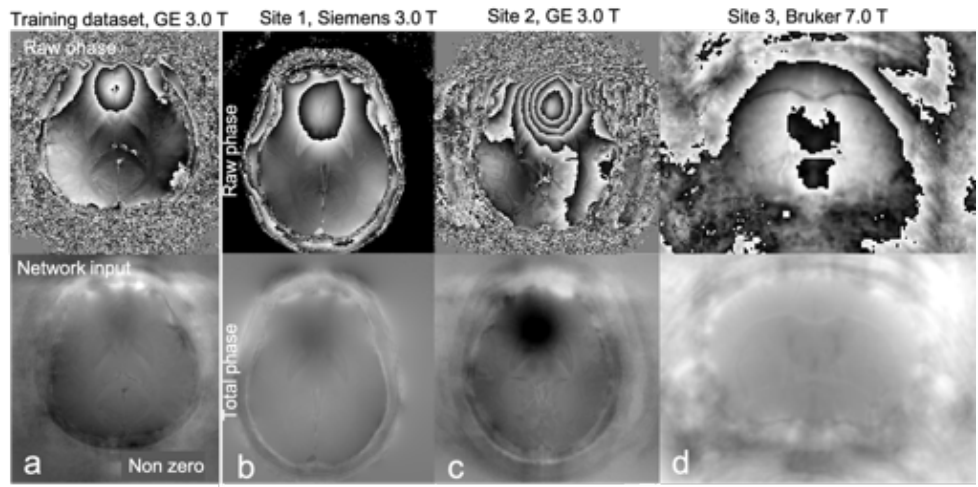

Fig. S6. Representative unwrapped phase images show different patterns outside of the brain acquired at different sites.

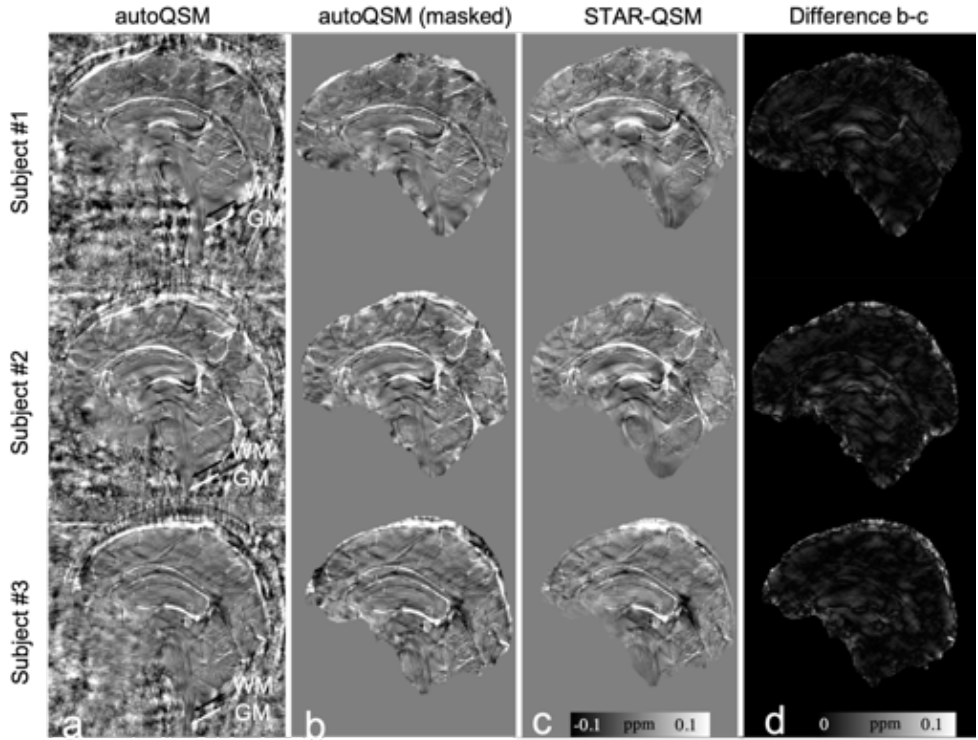

Fig. S7. Results of autoQSM for imaging the cortical vessels on a high-resolution dataset. (a); autoQSM's results; (b) masked autoQSM's results using the mask derived from STAR-QSM images; (c) STAR-QSM's results; (d) difference maps between b and c.

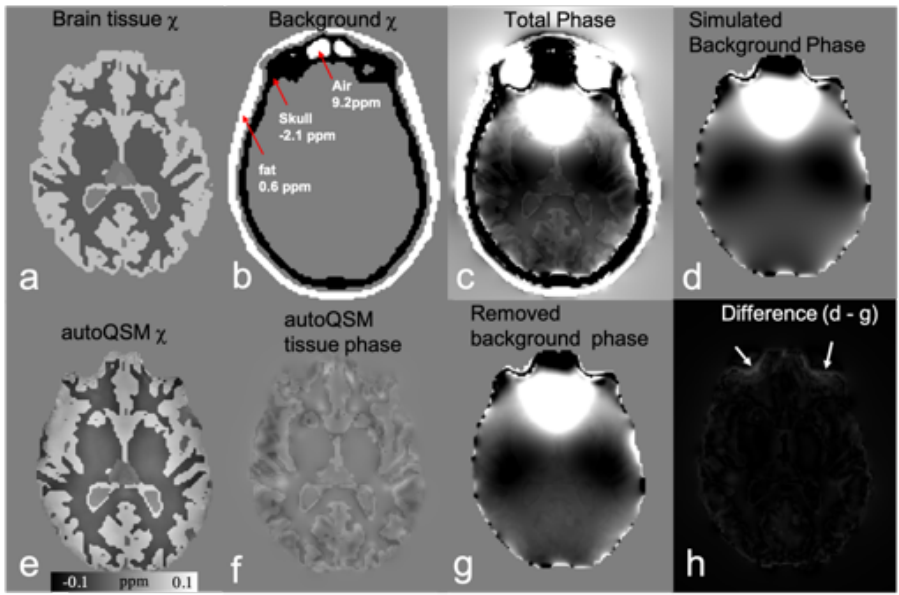

Fig. S8. The numerical brain phantom experiment to test autoQSM's performance for background phase removal.